\documentclass[preprint]{aastex}

\usepackage{amsmath}
\usepackage{graphicx}
\usepackage{hyperref}
\usepackage{subfig}
\usepackage[final]{pdfpages}
\usepackage{natbib}
\newcommand{\snr}{{\raise0.7ex\hbox{${\rm{S}}$} \!\mathord{\left/
 {\vphantom {{\rm{S}} {\rm{N}}}}\right.\kern-\nulldelimiterspace}
\!\lower0.7ex\hbox{${\rm{N}}$}}
}

\slugcomment{Original version appears as Chapter 2 in \textit{The Proceedings of SETI Sessions at the 2010 Astrobiology Science Conference: Communication with Extraterrestrial Intelligence (CETI), Douglas A. Vakoch, Editor}}

\begin{document}

\author{A. P. V. Siemion\altaffilmark{1}, J. Cobb\altaffilmark{1}, H. Chen\altaffilmark{2}, J. Cordes\altaffilmark{3}, T. Filiba\altaffilmark{1}, G. Foster\altaffilmark{4}, A. Fries\altaffilmark{1}, A. Howard\altaffilmark{1}, J. von Korff\altaffilmark{1}, E. Korpela\altaffilmark{1}, M. Lebofsky\altaffilmark{1}, P. L. McMahon\altaffilmark{5}, A. Parsons\altaffilmark{1},  L. Spitler\altaffilmark{3}, M. Wagner\altaffilmark{1} and D. Werthimer\altaffilmark{1}}

\email{siemion@berkeley.edu}

\altaffiltext{1}{University of California, Berkeley}
\altaffiltext{2}{University of California, Los Angeles, CA }
\altaffiltext{3}{Oxford University}
\altaffiltext{4}{Cornell University}
\altaffiltext{5}{Stanford University}

\pagenumbering{roman}
\title{Current and Nascent SETI Instruments}
\maketitle
\setcounter{page}{1}
\pagenumbering{arabic}

\section*{Abstract}
Here we describe our ongoing efforts to develop high-performance and sensitive instrumentation for use in the search for extra-terrestrial intelligence (SETI). These efforts include our recently deployed Search for Extraterrestrial Emissions from Nearby Developed Intelligent Populations Spectrometer (SERENDIP V.v) and two instruments currently under development; the Heterogeneous Radio SETI Spectrometer (HRSS) for SETI observations in the radio spectrum and the Optical SETI Fast Photometer (OSFP) for SETI observations in the optical band. We will discuss the basic SERENDIP V.v instrument design and initial analysis methodology, along with instrument architectures and observation strategies for OSFP and HRSS. In addition, we will demonstrate how these instruments may be built using low-cost, modular components and programmed and operated by students using common languages, e.g. ANSI C.

\section{Background}
By far the most common type of SETI experiments are searches for
narrowband continuous-wave radio signals originating from astronomical
sources. These searches are based on a number of fundamental principles,
many first described by Frank Drake in the early 1960s \citep{Drake:1961bv}. Paramount among them
is the fact that sufficiently narrow signals are easily distinguishable from
astrophysical phenomena, and would thus be a reasonable choice for a deliberate beacon from an advanced intelligence. The spectrally narrowest known astrophysical sources
of electromagnetic emission are masers, with a minimum frequency spread of
about one kHz. Additional support for the possible preference for narrowband
interstellar transmissions by ETI include the immunity of narrowband signals
to astrophysical dispersion and consideration of similarities to our own
terrestrial radio communication systems. Further encouragement is provided
by the existence of a computationally efficient matched filter for searching
for narrowband signals, the Fast Fourier Transform. More recently, searches
have begun targeting other signal types, such as broadband dispersed radio
pulses \citep{Siemion:2010p6845}.

Optical SETI, the name usually ascribed collectively to SETI operating at optical wavelengths, was first proposed in 1961 by Schwartz and Townes \citep{1961Natur.190..205S} shortly after the development of the laser. Searches have been conducted for both pulsed emission, e.g. \cite{2004ApJ...613.1270H}, and continuous narrowband lasers, e.g. \cite{2002PASP..114..416R}. Pulsed optical SETI rests on the observation that humanity could build a pulsed optical transmitter (using, for example, a U.S. National Ignition Facility-like laser and a Keck Telescope-like optical beam former) that could be detectable at interstellar distances. When detected, the nanosecond-long pulses would be a factor of $\sim$1000 brighter than the host star of the transmitter during their brief flashes \citep{2004ApJ...613.1270H}. Such nanosecond-scale optical pulses are not known to occur naturally from any astronomical source \citep{Howard:2001to}.

\subsection{Extant SETI Searches}

Our group is involved in a variety of ongoing searches for signatures of extraterrestrial intelligence, spanning the electromagnetic spectrum from radio to optical wavelengths. The most publicly well known of these is our distributed computing effort, SETI@home \citep{Anderson:2002p3265}.  Launched in 1999, SETI@home has engaged over 5 million people in 226 countries in a commensal sky survey for narrow band and pulsed radio signals near 1420 MHz using the Arecibo radio telescope.  SETI@home is currently operating over a 2.5 MHz band on the seven beam Arecibo L-band Feed Array (ALFA).  Participants in the project are generating the collective equivalent of 200 TeraFLOPs/sec and have performed over  $1.4 \times 10^{22}$ FLOPs to date. 

Another of our radio SETI projects, the Search for Extra-Terrestrial Radio Emissions from Nearby Developed Intelligent Populations (SERENDIP) \citep{Werthimer:1995p885}, is now in its fifth generation and is currently being conducted in a collaboration between UC Berkeley and Cornell University (Table \ref{tab:history}). In June 2009 we commissioned SERENDIP V.v, the newest iteration of the three-decade old SERENDIP program\footnote{SERENDIP experiment numbering proceeds from IV (four) to V.v (five point five) to accommodate the ambiguous naming of the SERENDIP V computing board, which was in fact never used for searches for extra-terrestrial intelligence.}. This project utilizes a high performance field programmable gate array (FPGA)-based spectrometer attached to the Arecibo ALFA receiver to perform a high sensitivity sky survey for narrow-band signals in a 300 MHz band surrounding 1420 MHz. The SERENDIP V.v spectrometer analyzes time-multiplexed signals from all seven dual-polarization ALFA beams, commensally with other telescope users, effectively observing 2 billion channels across seven 3 arc-minute pixels. A copy of this instrument is currently deployed by the Jet Propulsion Laboratory on a 34-m Deep Space Network (DSN) dish, DSS-13, in Barstow, California.
\begin{table}[tb]
\footnotesize
\caption{HRSS Costs Compared To Other SETI Spectrometers}
\centering
\vspace{0.1in}
\begin{tabular}{lccccc}
\hline\hline\\[-0.08in]
\textbf{Program} & \textbf{Bandwidth} & \textbf{Resolution} & \textbf{Channels} & \textbf{Date} & \textbf{Location} \\
  & (MHz) & (Hz) &   &  &  \\[0.03in]
\hline\\[-0.10in]
SERENDIP I  & 0.1 & 1000 & 100 & 1979-1982 & Hat Creek, Goldstone \\
\hspace{0.2in} \\[0.1in]
SERENDIP II & 0.065 & 1 & 64K & 1986-1990 & Green Bank, Arecibo \\
\hspace{0.2in} \\[0.1in]
SERENDIP III  & 12 & 0.6 & 4M & 1992-1996 & Arecibo \\
\hspace{0.2in} \\[0.1in]
SERENDIP IV  & 100 & 0.6 & 168M & 1998-2006 & Arecibo \\
\hspace{0.2in} \\[0.1in]
SERENDIP V.v  & 300 & 1.5  & 2G$^1$ & 2009 - & Arecibo\\
\hline
\end{tabular}
\label{tab:history}
\footnotetext[1]{table footnote 1}
\begin{flushleft} $^1$ SERENDIP V.v is currently multiplexed. \end{flushleft}
\end{table}

\begin{figure}[tb]
	\centering
		\includegraphics[width=0.9\linewidth]{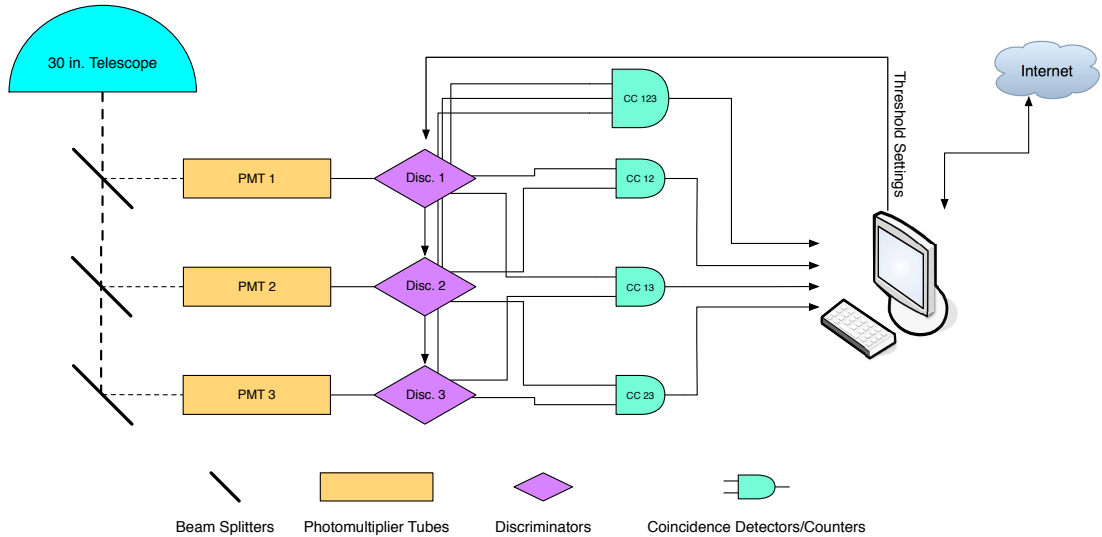}
		\caption{\footnotesize{Schematic diagram for the existing analog electronics and ISA digital counter card used to read out the Lick and Leuschner SETI photometers. An optical telescope feeds three photomultiplier tubes using optical beamsplitters. Analog electronics threshold the signals from three PMTs and coincidence detectors trigger low-speed counters. While this system is effective all detailed data about the event trigger (the digitized light profile) is lost.}}
	\label{fig:pmtplot_old}
\end{figure}

Our optical pulse search \citep{2000ASPC..213..565L} is based at UC Berkeley's 30-inch automated telescope at Leuschner Observatory in Lafayette, California. The detector system consists of a custom-built photometer, employing three photomultiplier tubes (PMTs) fed by an optical beamsplitter to detect the concurrent (within  $\sim$1 ns) arrival of incoming photons across a wavelength range $\lambda = 300$--650 nm. This ``coincidence'' detection technique improves detection sensitivity by reducing the false alarm rate from spurious and infrequent pulses observed in individual PMTs.   PMT signals are fed to three high speed amplifiers, three fast discriminators and a coincidence detector (Figure \ref{fig:pmtplot_old}), where detections are measured by a relatively slow (1 MHz) Industry Standard Architecture (ISA) counter card. The photometer features a digitally adjustable threshold level to set the false alarm rate for a particular sky/star brightness. 
During a typical observation, the telescope is centered on a star and detection thresholds are adjusted so that the false alarm rate is sufficiently low. Currently we record three types of events: single events, when an individual PMT output is greater than the voltage threshold originally set; double events, when any two of the PMTs output exceeds the threshold in the same nanosecond-scale time period; and triple events, when all three PMTs concurrently exceed threshold. Voltage thresholds are set so that false triple events are very rare and false double events occur only a few times in a 5 minute observation.  A duplicate of this instrument is in place at Lick Observatory near San Jose, California \citep{2005AsBio...5..604S}.

\subsection{New Instruments}
Historically, the level of technology and engineering expertise required to implement a SETI instrument was quite high. As a result, SETI programs have been limited to just a handful of institutions. Our group is developing two new instruments possessing several advantages over previous generations of SETI instrumentation -- an Optical SETI Fast Photometer for optical SETI and the Heterogeneous Radio SETI Spectrometer for observations in the radio. Both are constructed from widely available modular components with relatively simple interconnects. The designer logs into the instrument using Linux and programs in C, obviating the need for cumbersome interfaces (e.g. JTAG) and languages (e.g. VHDL/Verilog). Further, these instruments are scalable and easily upgradable by adding additional copies of commercially available parts (compared with the money and time-consuming upgrades of previous instruments that involved complete redesigns of PC boards and ASICs). Collectively these advances will enable much wider participation in SETI science.  

Our next generation Optical SETI Fast Photometer (OSFP) is based on the same front-end optics and photodetectors as the original Berkeley OSETI instrument, but adds a flexible digital back-end based on the Center for Astronomy Signal Processing and Electronics Research (CASPER, see below) DSP instrument design system. The programmable FPGA-based digital back-end will allow us to improve sensitivity by implementing sophisticated real-time detection algorithms, capture large swaths of raw sampled voltages for diagnostics or centroiding and perform efficient rejection of interference based on pulse profiles.

Our newest radio SETI instrument, the Heterogeneous Radio SETI Spectrometer (HRSS), is also CASPER-based.  HRSS will take advantage of the wide bandwidth capabilities of a high-speed analog-to-digital converter (ADC) paired with an FPGA to digitize, packetize, and transmit coarse channelized spectral regions to flexible, off-the-shelf CPUs and graphics processing units (GPUs) for fine spectroscopy and RFI rejection. This architecture will not only provide for economical entry into cutting edge SETI research (Table \ref{tab:costs}, below), its use of standard C programming on CPUs and GPUs will enable the DSP instrument internals to be accessible for students with only modest instrumentation experience. The HRSS architecture is highly scalable and inexpensive, paving the way for future spectrometers with very large bandwidths (many GHz) covering many beams simultaneously. 

The complete instrument system for both HRSS and OSFP, including digitization and packetization hardware, digital signal processing (DSP) algorithms and control software, will be made publicly available for students and researchers worldwide. 
\subsection{Open Source Hardware Infrastructure}
All of the instruments discussed here take advantage of the open source, modular DSP instrumentation framework developed by the Center for Astronomy Signal Processing and Electronics Research (CASPER) \citep{Werthimer:2011p9714}. This international collaboration seeks to shorten the astronomy instrument development cycle by designing modular, upgradeable hardware and a generalized, scalable architecture for combining this hardware into a signal-processing instrument. Employing FPGAs, FPGA-based chip-independent signal processing libraries, and packetized data routed through commercially available switches, CASPER instrument architectures look like a Beowulf cluster, with reconfigurable, modular computing hardware in place of CPU compute nodes. Thusly, a small number of easily replaceable and upgradeable hardware modules may be connected with as many identical modules as necessary to meet the computational requirements of an application, known colloquially as ``computing by the yard.''  Such an architecture can provide orders of magnitude reduction in overall cost and design time and closely tracks the early adoption of state-of-the-art IC fabrication by FPGA vendors.

The Berkeley Emulation Engine (BEE2) system was CASPER's first attempt at providing a scalable, modular, economic solution for high-performance DSP applications (Chang et al., 2005). The BEE2 system consists of three hardware modules: the main BEE2 processing board, a high-speed ADC board for data digitization and an iBOB board primarily responsible for packetizing ADC data onto the Ethernet protocol. Communication between hardware modules takes place over standard 10 Gbit Ethernet (10 GbE) links, allowing for the relatively simple integration of commercial switches and processors.

The current generation Virtex-5-based ÒROACHÓ board (Reconfigurable Open Architecture for Computing Hardware) replaces, but interoperates with, both the BEE2 and IBOB boards. ROACH includes a single Xilinx Virtex-5 FPGA (SX95T, LX110T, LX155T), four 10 GbE-CX4 ports, 2 ADC ports, up to 8GB of DDR2 memory, 72Mbit of QDR and an independent control and monitoring PowerPC processor. ROACH remains compatible with all current and next generation ADC boards.

All CASPER boards may be programmed via a set of open-source libraries for the Simulink/Xilinx System Generator FPGA programming language. These libraries abstract chip-specific components to provide high-level interfaces targeting a wide variety of devices. Signal processing blocks in these libraries, such as polyphase filterbanks, Fast Fourier Transforms, digital down converters and vector accumulators, are parameterized to scale up and down to arbitrary sizes, and to have selectable bit widths, latencies and scaling.

\section{SERENDIP V.v}
Over the last 30 years SERENDIP spectrometer development has closely tracked the Moore's Law growth in the electronics industry, with new spectrometers processing ever-larger bandwidths while achieving finer spectral resolution. SERENDIP V.v is the most powerful spectrometer yet built as part of the SERENDIP project. SERENDIP V.v was installed at Arecibo Observatory in June 2009 and operates commensally with other experiments on the ALFA multi-beam receiver. Currently, the spectrometer multiplexes beam-polarizations through a single-beam 200MHz digital signal processing chain via a computer-controlled RF switch. 

\begin{figure}[tb]
	\centering
		\includegraphics[width=0.9\linewidth]{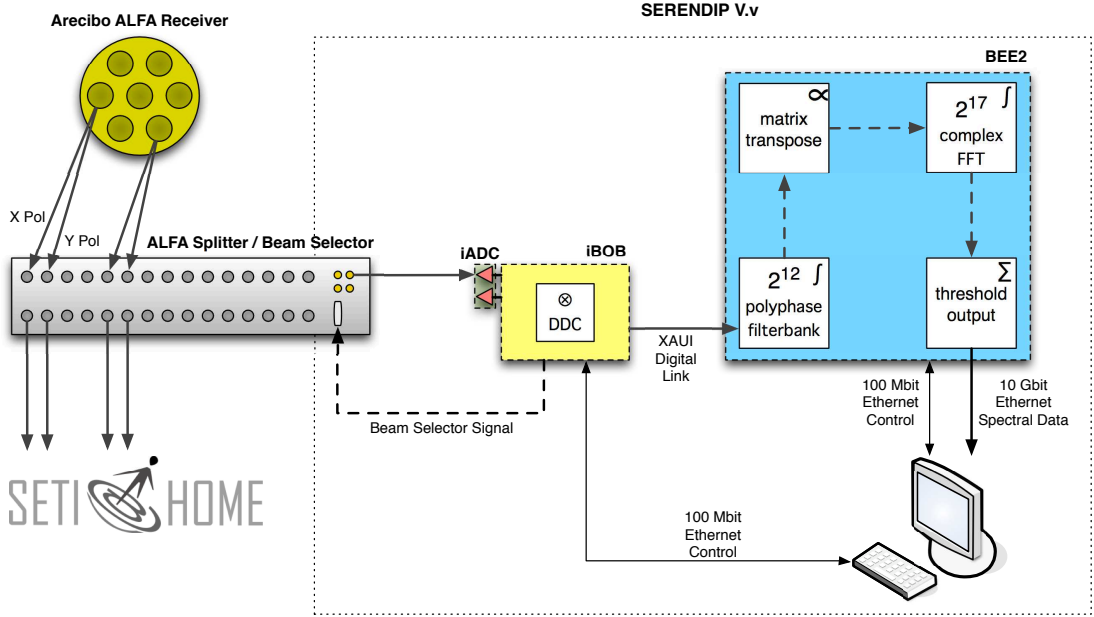}
		\caption{\footnotesize{SERENDIP V.v instrument architecture.  Analog signals from the ALFA receiver, mixed
down to IF, are fed to a computer-controlled switch.  One copy of the input is relayed to the SETI@home data recorder and a time-multiplexed beam is sent to the SERENDIP V.v spectrometer.  The spectrometer samples the incoming IF signal at 800 Msamples/sec, digitally down converts the data to a complex baseband representation, performs a two-stage channelization (yielding $\sim$1 Hz spectral resolution) and outputs over-threshold frequency
channels to a host PC.}}
	\label{fig:serendip}
\end{figure}

The SERENDIP V.v system architecture and dataflow are shown in Figure \ref{fig:serendip}.  ALFA signals for all 14 beam-polarizations are fed into an RF switch, with a single output fed into a high-speed ADC sampling at 800 Msps. An iBOB board mixes the sampled signal down to baseband, decimates to a 200 MHz bandwidth and transmits the serialized data stream to a BEE2 via a high-speed digital link. Processing on the BEE2 is split into four stages, each of which occupies a separate FPGA on the board. The data stream is 1: coarse channelized via a 4096pt polyphase filter bank (PFB),  2: matrix transposed by a Ôcorner turnerÕ to facilitate a second stage of channelization, 3: fine channelized using a conventional 32768pt Fast Fourier Transform (FFT) and finally 4: Ôthresholded,Õ in which each fine frequency ÔbinÕ (1.49 Hz wide) is compared against a scaled coarse-bin average to pick out fine bins of interest. Local averages are calculated per PFB channel by averaging the same data being fed to the FFT in parallel with the transform.  This way, the total power in each PFB bin can be accumulated while the FFT is being computed (via Parseval's theorem). 

The thresholding process triggers ``hits'' for fine/FFT bins that are greater than or equal to the threshold power. For practical reasons, the number of hits reported per coarse/PFB bin is capped via a software-adjustable setting, usually set to report fine bins between 15-30 times the average power. The reported hits are assembled into UDP packets on-board the BEE2 and transmitted to a host PC. The host PC combines spectrometer data with meta-information, such as local oscillator settings and pointing information, and writes the complete science data stream to disk. To-date, SERENDIP V.v has commensally observed for approximately 900 hours. Analysis efforts are underway, in parallel, at both UC Berkeley and Cornell. 

While both SERENDIP V.v and SETI@home operate simultaneously and commensally on the same RF signal, SERENDIP V.v differs in the key respect that the computationally intensive Fourier Transform is performed internally, rather than through distributed computing. This forces the SERENDIP V.v spectrometer to use a much simpler search algorithm than SETI@home employs. However, since the SERENDIP spectrometer is collocated with the telescope, it has access to a much larger bandwidth. SERENDIP and SETI@home are thus complementary, in that together they can look with both a panoramic gaze across many MHz and with microscopic precision near the 21cm ``watering hole.''

\section{Heterogenous Radio SETI Spectrometer}
The HRSS instrument system bridges our previous radio SETI programs by connecting open source FPGA-based signal processing hardware and software to an easily-programmable GPU-equipped multicore CPU back-end, thus achieving an economical student-friendly SETI instrument. The low cost, scalable architecture used in HRSS will enable more widespread deployment than previous instruments, potentially increasing both the sky and frequency coverage of the radio SETI search space. With previous instruments, difficulty in programming the hardware precluded implementing intricate algorithms directly into the real-time data flow. The flexibility of the CPU/GPU back-end of HRSS will readily enable arbitrarily sophisticated algorithms in the real-time processing pipeline, including dynamic interference rejection and immediate follow-up.

\begin{figure}[tb]
\begin{center}
\begin{tabular}{c}
\includegraphics[width=1.05\linewidth]{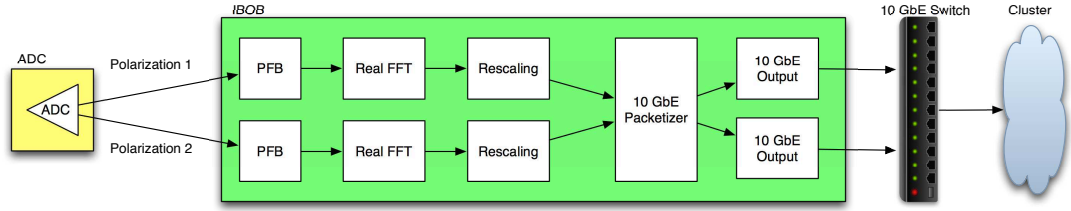}
\end{tabular}
\vspace{-3.0in} %figure has lots of whitespace at the bottom
\caption{\footnotesize{The Packetized Astronomy Signal Processor Block Diagram: the data flow inside the FPGA. Two data streams from the ADC are channelized by the PFBs and FFTs. After rescaling and bit selection (18 bits down to 8 bits), the data are packetized and sent out over two 10GbE links to a server or 10GbE switch.}}
\label{fig:packet_spec}
\end{center}
\end{figure}

The prototype for HRSS is the existing Packetized Astronomy Signal Processor (PASP) \citep{McMahon:2011tk}, based on the CASPER iBOB. This reconfigurable FPGA design channelizes two signals, each of 400 MHz bandwidth (digitizing at 800 Msps), packetizes, and distributes the channels to different IP addresses using a runtime programmable schema. Figure \ref{fig:packet_spec} shows a block diagram of the PASP instrument. Two signals (e.g. two polarizations) are fed into an iBOB using a dual ADC board. Each polarization is sent through a PFB, which channelizes the streams. Individual channels are buffered into packets and sent out over 10 GbE links to a cluster of servers via a 10 GbE switch or a single backend server directly connected to the iBOB. 

The PASP design is highly reconfigurable. The number of channels, number of IP addresses, and the packet size can all be easily adjusted in the instruments Simulink design. This design can support a variety of back-end processing options simply by adjusting these three parameters. The number of channels adjusts the size of the sub-band for each processing element. Dividing the 400 MHz band into 16 channels creates large 25 MHz sub-bands, which may require a faster server, but this can be balanced by increasing the number of channels and thereby reducing the size of the sub-bands and processing demand. The number of IP addresses also controls the bandwidth each backend server receives. In a 16-channel design with only 8 IP addresses, each IP will receive 2 channels. In a server with multiple processing elements (e.g. multiple CPU cores or GPUs), these channels can be processed in parallel.  The FPGA portion of HRSS largely consists of a port of the PASP design to the new CASPER ROACH board, taking advantage of the larger FPGA, enabling a larger bandwidth and improved interface. The iBOB can be difficult to interface with, requiring a JTAG connection to reprogram the board and a very limited shell program to interact with the FPGA. In contrast, ROACH provides a full Linux OS.

Additional software running on connected CPUs/GPUs will finely channelize the sub-bands and identify possible events for further processing. This software will initially be developed in ANSI-C to allow maximum portability. Once the C-based system is fully prototyped, we will optimize for GPU hardware and specialized languages to extract more processing power from servers with graphics capabilities. We are investigating both OpenCL and CUDA as target languages. CUDA will provide excellent performance, but can only be compiled for NVIDIA GPUs.  OpenCL is designed to compile for generic CPU and GPU platforms, but it may not provide the performance efficiency of an architecture-specific language like CUDA.

\begin{figure}[tb]
\begin{center}
\begin{tabular}{c}
\includegraphics[width=0.8\linewidth]{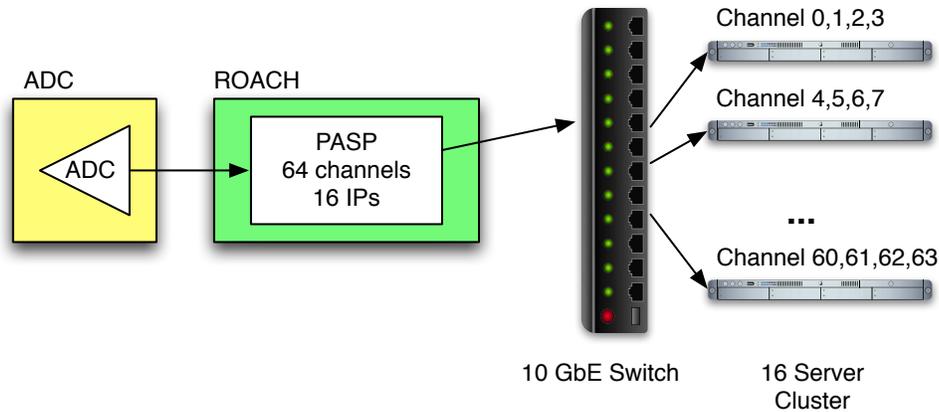}
\end{tabular}
\caption{\footnotesize{The Heterogeneous Radio SETI Spectrometer Block Diagram: an example configuration of the HRSS instrument. The ROACH channelizes ADC data into 64 channels and sends it over a 10GbE switch to a cluster of 16 servers. The number of channels and servers can be reconfigured based on available server processing power.}}
\label{fig:packet_spec_high_level}
\end{center}
\end{figure}

Figure \ref{fig:packet_spec_high_level} shows an example configuration of HRSS with a cluster of servers on the backend. The figure shows a PASP configured for 64 channels and 16 IPs, a 10GbE switch, and a cluster of backend servers. The reconfigurability of the PASP design makes the required size and computing power of the backend processing cluster highly elastic, scaling from a single server to a cluster of high-powered servers.

As shown in Table \ref{tab:costs}, HRSS is extremely cost effective compared to other SETI spectrometers with $\sim$1 Hz spectral resolution. HRSS is less expensive than SERENDIP V.v, primarily because it uses a newer single-FPGA ROACH board paired with commodity computing hardware instead of an iBOB with a 5-FPGA BEE2 board. The HRSS architecture can be easily scaled up to process a 1.5 GHz dual polarization signal on a single ROACH board using currently available dual 3 Gsps ADCs and multiple fine channelization nodes. The cost of each additional 125 MHz/dual polarization module is about a factor of three less than the first module.

\begin{table}[tb]
\footnotesize
\caption{HRSS Costs Compared To Other SETI Spectrometers}
\centering
\vspace{0.1in}
\begin{tabular}{lccccc}
\hline\hline\\[-0.08in]
\textbf{SETI Spectrometer} & \textbf{Bandwidth} & \textbf{Beams} & \textbf{Pol's} & \textbf{Cost$^1$} & \textbf{Normalized Cost} \\
  &  &   &   &  & per MHz/beam/pol \\[0.03in]
\hline\\[-0.10in]
SERENDIP V.v  & &  &  &  & \\
\hspace{0.2in} UCB, deployed at Arecibo \& JPL & 200 MHz & 1 & 1 & \$40K & \$200\\[0.1in]
HRSS & &  &  &  & \\
\hspace{0.2in} first 125 MHz dual pol bands & 125 MHz & 1 & 2 & \$9K & \$40\\[0.1in]
HRSS  & &  &  &  & \\
\hspace{0.2in} additional 125 MHz dual pol bands & 125 MHz & 1 & 2 & \$3K & \$15\\
\hspace{0.2in} (after the first 125 MHz) &  &  & &  & \\[0.05in]
\hline
\end{tabular}
\label{tab:costs}
\footnotetext[1]{table footnote 1}
\begin{flushleft} $^1$ Costs do not include labor. \end{flushleft}
\end{table}

\section{Optical SETI Fast Photometer}
The forthcoming Optical SETI Fast Photometer (OSFP) is based on the same front-end optics and photodetectors as the original Berkeley OSETI instrument, but adds a flexible digital back-end based on the CASPER DSP instrument design system. This instrument will significantly improve our sensitivity to pulsed optical signals, and lower some of the barriers to wider engagement in optical SETI searches.
The digital back-end for the instrument, Figure 5, will be constructed from modular CASPER components; direct sampling PMT outputs with two dual 8-bit, 1500 Msps ADC boards and using a single ROACH board for DSP. This board features a variety of interfaces for connection to a control computer, accommodating a variety of experiment parameters. For high threshold, low event rate searches, the ROACH's 100 Mbit Ethernet should be sufficient for data acquisition. For low thresholds, or characterization of instrument PMTs, the ROACH's 10GbE interfaces can be used for transferring many events and/or large swaths of raw sampled voltages.

\begin{figure}[tb]
	\centering
		\includegraphics[width=0.9\linewidth]{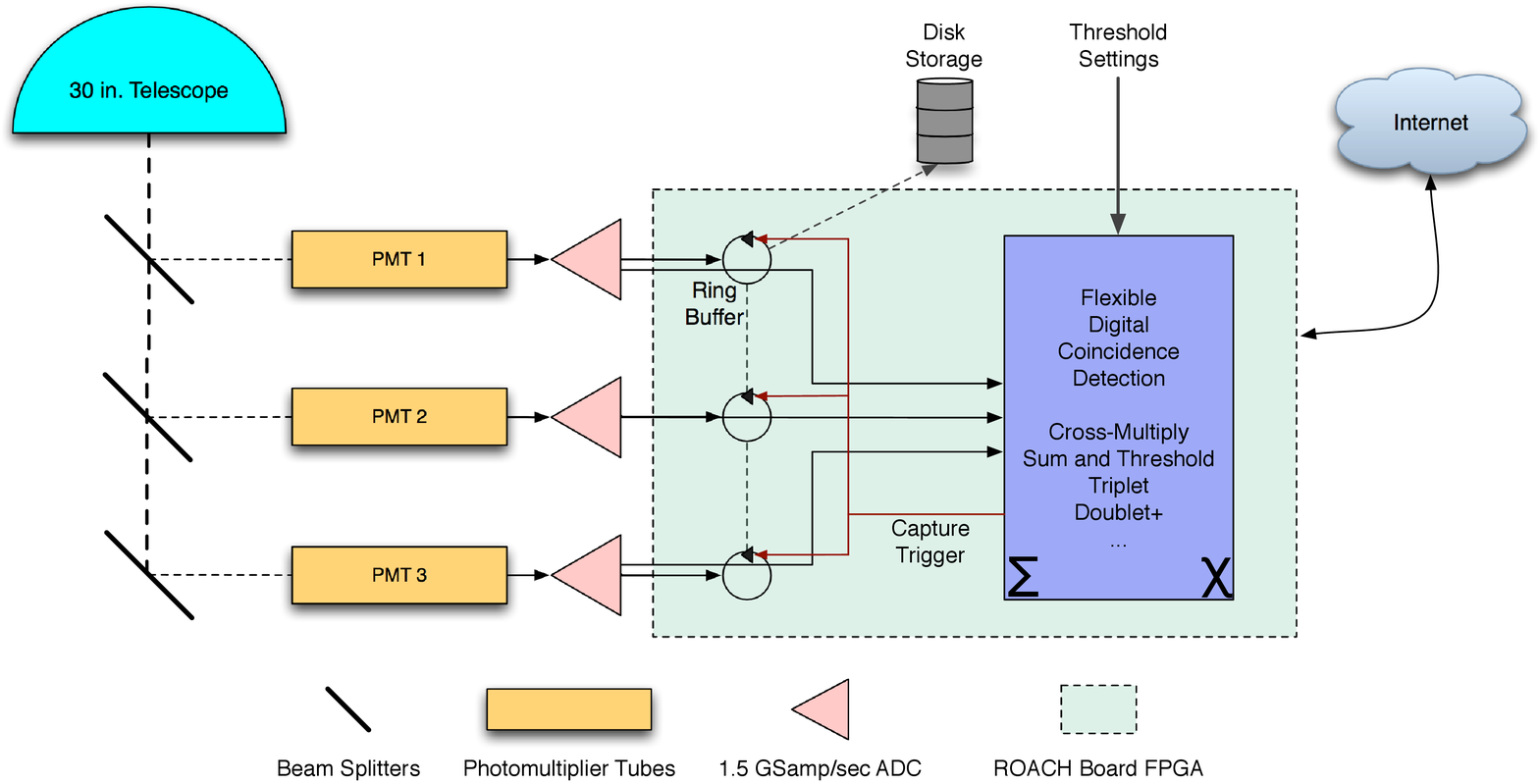}
		\caption{\footnotesize{Schematic diagram for the OSFP digital backend. An optical telescope feeds three photomultiplier tubes (PMTs) using optical beamsplitters. The PMT outputs are digitized directly using 1.5 Gsamp/sec ADCs and transmitted to a ROACH board for processing. Onboard the ROACH, voltage samples are copied into a 4 Gb DRAM ring buffer that feeds a  programmable event detection circuit.  When triggered, the ring buffer contents are read out to a host computer, capturing raw event data. The digital logic for the instrument is fully compatible with the CASPER open-source instrument design tool flow.}}
	\label{fig:pmtplot_updated}
\end{figure}

The programmable FPGA-based digital back-end will allow us to improve sensitivity by implementing sophisticated real-time detection algorithms. In our existing system at Leuschner observatory, the detection algorithm is very simple -- all three PMT signals must be above a programmable threshold to trigger an event. With OSFP, one can implement more sophisticated detection algorithms. For example, a multistage trigger could be implemented that requires the sum of the three digitized PMT outputs to exceed a threshold as well as the requirement that the signal levels in the three streams be similar to each other.  The ability to perform significant computations on the data streams in real time is a crucial aspect of this design. We envision searching for multiple signal types simultaneously, including weak pulse trains with repetition times from ns to ms and violations of Poisson statistics in photon arrival times (indicating a non-astrophysical source). False positive signals can also be efficiently rejected based on pulse profiles (a capability sorely lacking in the current threshold-based instrument).

The large amount of DRAM available on the ROACH board will enable buffering of raw PMT waveforms and triggered write-to-disk based on high-confidence events. Such a capability will enable detailed analysis of an event, including precise determination of pulse arrival times using centroiding. Upon detection of a coincidence event, a user-adjustable section of the corresponding waveform, along with microsecond time-tagging provided by a GPS 1 pulse per-second system, will be packetized and transmitted to a host computer over one of the ROACH's Ethernet interfaces. A parallel, streaming DSP design will enable the instrument to operate at 100\% duty with reasonable waveform buffers. Should a significant event be detected, the software system will automatically alert the observer to the possible signal detection, and will optionally automatically cause the telescope to continue observing the same sky coordinates where the telescope was pointed when the reported flash arrived.

In anticipation of the real-time computing capabilities of OSFP, we have performed preliminary simulations of several pulse detection algorithms. These simulations model the entire front-end of the system, including the optical beam splitter, PMTs, and ADCs.  In initial work, it appears the optimal algorithm involves thresholding the cross-correlation of each pair of PMT waveforms, but we continue to evaluate tradeoffs in sensitivity and false alarm rate.  Future simulations will allow us to improve our algorithms by incorporating more elaborate detection criteria.

Use of CASPER hardware and gateware for this instrument guarantees an upgrade path when faster ADCs become available, eventually allowing full Nyquist sampling of the PMT bandwidth. All optics and detector components for the existing front-end are available off-the-shelf from Hamamatsu and Edmunds Industrial Optics. The entire assembly can be constructed without special tools, and complete instructions and parts lists are available at \url{http://seti.berkeley.edu/opticalseti}.

%% In this section, we use  the \subsection command to set off
%% a subsection.  \footnote is used to insert a footnote to the text.


%% This section also includes several of the displayed math environments
%% mentioned in the Author Guide.

\section{Acknowledgments}

The Berkeley SETI projects are funded by grants from NASA and the National Science Foundation, and by donations from the friends of SETI@home. We acknowledge generous donations of technical equipment and tools from Xilinx, Fujitsu, Hewlett Packard and Sun Microsystems.


%% in parentheses, after the keyword, but they will not be verified.

{\it Facilities:} \facility{Arecibo}, \facility{Leuschner}.

\clearpage

\bibliography{references}

\begin{thebibliography}{12}
\expandafter\ifx\csname natexlab\endcsname\relax\def\natexlab#1{#1}\fi

\bibitem[{Anderson {et~al.}(2002)Anderson, Cobb, Korpela, Lebofsky, \&
  Werthimer}]{Anderson:2002p3265}
Anderson, D.~P., Cobb, J., Korpela, E., Lebofsky, M., \& Werthimer, D. 2002,
  Communications of the ACM

\bibitem[{Drake(1961)}]{Drake:1961bv}
Drake, F.~D. 1961, Physics Today, 14, 40

\bibitem[{Howard \& Horowitz(2001)}]{Howard:2001to}
Howard, A., \& Horowitz, P. 2001, Proc. SPIE, 4273

\bibitem[{Howard {et~al.}(2004)Howard, Horowitz, Wilkinson, Coldwell, Groth,
  Jarosik, Latham, Stefanik, Willman, Wolff, \& Zajac}]{2004ApJ...613.1270H}
Howard, A.~W., {et~al.} 2004, The Astrophysical Journal, 613, 1270

\bibitem[{Lampton(2000)}]{2000ASPC..213..565L}
Lampton, M. 2000, ASP Conference Series, 213, 565

\bibitem[{McMahon(2008)}]{McMahon:2011tk}
McMahon, P.~L. 2008, arXiv:1109.0416 [astro-ph.IM], M.Sc. Thesis, University of
  Cape Town

\bibitem[{Reines \& Marcy(2002)}]{2002PASP..114..416R}
Reines, A.~E., \& Marcy, G.~W. 2002, The Publications of the Astronomical
  Society of the Pacific, 114, 416

\bibitem[{Schwartz \& Townes(1961)}]{1961Natur.190..205S}
Schwartz, R.~N., \& Townes, C.~H. 1961, Nature, 190, 205

\bibitem[{Siemion {et~al.}(2010)Siemion, Korff, McMahon, Korpela, Werthimer,
  Anderson, Bower, Cobb, Foster, Lebofsky, Leeuwen, \&
  Wagner}]{Siemion:2010p6845}
Siemion, A., {et~al.} 2010, Acta Astronautica, 67, 1342

\bibitem[{Stone {et~al.}(2005)Stone, Wright, Drake, Mu{\~n}oz, Treffers, \&
  Werthimer}]{2005AsBio...5..604S}
Stone, R. P.~S., Wright, S.~A., Drake, F., Mu{\~n}oz, M., Treffers, R., \&
  Werthimer, D. 2005, Astrobiology, 5, 604

\bibitem[{Werthimer {et~al.}(2011)Werthimer, Backer, \&
  Wright}]{Werthimer:2011p9714}
Werthimer, D., Backer, D., \& Wright, M. 2011, \url{http://casper.berkeley.edu}

\bibitem[{Werthimer {et~al.}(1995)Werthimer, Ng, Bowyer, \&
  Donnelly}]{Werthimer:1995p885}
Werthimer, D., Ng, D., Bowyer, S., \& Donnelly, C. 1995, Astronomical Society
  of the Pacific Conference Series, 74, 293, iSBN: 0-937707-93-7

\end{thebibliography}
\appendix

\clearpage

\end{document}